\documentclass[%
 reprint,
 amsmath,amssymb,
 aps,
]{revtex4-2}

\usepackage{graphicx}
\usepackage{dcolumn}
\usepackage{bm}
\usepackage{hyperref}

\begin{document}

\preprint{APS/123-QED}

\title{Photon Calibration Techniques for High Resolution Cryogenic Detectors}

\author{W.~Matava}\email{wmatava@berkeley.edu}
\affiliation{University of California Berkeley, Department of Physics, Berkeley, CA 94720, USA}
\affiliation{Lawrence Berkeley National Laboratory (LBNL), Berkeley, CA 94720-8099, USA}

\author{M.R.~Williams}\email{michaelwilliams@lbl.gov}
\affiliation{Lawrence Berkeley National Laboratory (LBNL), Berkeley, CA 94720-8099, USA}

\begin{abstract}
Monoenergetic photons from a pulsed laser diode or LED are commonly used to calibrate the detector response of high-resolution calorimetric detectors. However, when the detector's resolution is larger than the energy of a single photon, a calibration is normally derived using Poisson statistics. In this paper, we clarify the assumptions implicit in this calibration method, before considering how a more realistic model of a detector's performance will violate these assumptions, biasing the calibration. Finally, we judge the individual impact of specific detector parameters on our calibration, and conclude with discussion of both the limits of our calculations and the implications for state-of-the art detectors.
\end{abstract}

\maketitle


\section{\label{sec:level1}Introduction}

Cryogenic solid-state detectors such as Transition Edge Sensors (TESs) and Kinetic Inductance Detectors (KIDs) have been widely used for bolometric measurement of photon power from cosmological sources \cite{SouthPoleTES}, celestial bodies \cite{HabitableWorlds}, as well as calorimetric measurement of astrophysical x-rays and gamma rays \cite{ATHENA_TES}. Searches for rare low-energy phenomenon such as low-mass dark matter recoils and coherent neutrino-nucleus scattering have motivated the development of superconducting phonon-mediated calorimetric detectors. In such application, the detectors are optimized for sensitivity to energy deposited in the substrate on which they are instrumented rather than directly into the sensing element. While TESs have demonstrated sub-eV resolution to phonon mediated events \cite{TwoChannelLimits,CDMS_HVeV_2026}, more emergent technologies, such as qubit-based sensors~\cite{qpd2024,SQUATPaper} and phonon-sensitive KIDs, have yet to reach such sensitivities \cite{TemplesKID,BullKID_orig,Calder_Final}.

While these detectors all exploit different mechanisms for sensing, they share a common mechanism for phonon collection. Incident radiation is absorbed in the detector substrate, creating a population of athermal phonons of number proportional to the deposited energy. These phonons propagate through the substrate as they anharmonically decay into lower energy phonons and scatter off of defects. After a number of reflections and scatters, they may reach the sensitive area of the detector. Here, phonons with energy greater than twice the superconducting band-gap of the material ($2\Delta_{SC}$) will break Cooper pairs and create a population of Bogoliubov quasiparticles (i.e. broken Cooper pairs) \cite{irwinQuasiparticleTrapAssisted1995}. These quasiparticles will either thermalize and change the temperature of the film in the case of TESs, change the kinetic inductance of the film in the case of KIDs, or tunnel across a Josephson junction in the case of qubits. These changes are then measured as the signal from the interaction. Historically, some technologies have aimed to employ "collection fins" made of a fully superconducting material with a band-gap such that $\Delta_{fin} \gg \Delta_{sensor}$. These different films generally overlap with the actual sensing element, allowing quasiparticles to diffuse across the band-gap gradient into the sensing element, where they can become trapped after downconversion. The fraction of the energy initially deposited into the substrate that arrives in the sensing element is hereafter referred to as the \textit{energy collection efficiency}. To date, only the TES technology has been successful in implementing this technique (dubbed the QET) \cite{irwinQuasiparticleTrapAssisted1995}.

Such high-resolution detectors can have correspondingly low saturation energies, potentially precluding their calibration using emissions from common radioisotopes~\cite{WilliamsFastNeutron}. Moreover, low-energy $\mathcal{O}$(keV) x-rays that may not cause saturation usually cannot penetrate the nested IR shielding characteristic of the low-temperature cryostats in which such detectors must be housed, requiring that radioactive material be placed in the cryostat alongside the detector. The constant presence of radioactive events can be a non-starter for experiments requiring radiopure environments, and may otherwise interfere with studying instrumental backgrounds, requiring multiple cool-downs to fully characterize a detector.

An alternative method is to use a controllable photon source such as a laser or LED \cite{LANTERN, TwoChannelPaper, CDMS_HVeV_2026}. Photon sources ranging from 375 nm (3.3 eV) to 11 $\mu$m (0.11 eV) are easily accessible to most experiments, and can be used without saturating most detectors. In most setups, photons are generated with a light source at room temperature and are transmitted into the cryostat using an optical fiber, which must be adequately thermalized and potentially filtered to prevent blackbody IR photons from leaking into the sample space and degrading detector performance through parasitic heating. In addition, these sources' intensities can be tuned to allow for the sensing of single photons \cite{TwoChannelPaper}. 

\begin{figure}[b]
\includegraphics[width=\columnwidth]{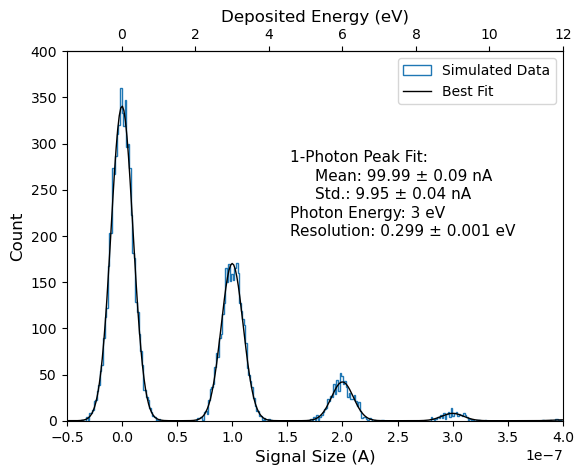}
\caption{\label{fig:subphot_res} Simulated histogram of responses from a TES with sub-eV resolution to a pulsed 3 eV light source. The number of photons being absorbed in each pulse can be resolved, yielding multiple known points of correspondence between the detector's response (measured in amps) and the amount of energy deposited. Using a fit to the 1-photon peak, (and the energy of photons being emitted), we can calibrate the detector's energy scale and quantify its resolution.}
\end{figure}

Calibration of a detector with resolutions sufficient to sense single photons is trivial, as shown in Fig. \ref{fig:subphot_res} \cite{TwoChannelLimits}. When the light source is pulsed, the difference in N and N+1 photons hitting the substrate can be resolved, and a histogram of pulse amplitudes can be fit to a sum of gaussians, with the means defining the calibration constant and the standard deviations defining the resolution.

However, emerging detector technologies often have resolutions much larger than any available pulsed photon source\cite{Calder_Final,TemplesKID,BullKID_orig}, potentially due to either poor intrinsic sensor resolution or poor energy collection efficiencies. It is for such detectors that calibrating using pulsed photons becomes more subtle, relying on statistical arguments rather than direct observation of energy peaks.

We will begin by reproducing the statistical arguments commonly used to calibrate sensors with greater-than-single-photon resolution, paying special attention to the underlying assumptions made. We will then present a more realistic model of the physics underlying such calibrations, and quantify the systematics stemming therefrom. After considering the magnitude of systematics in a typical sensor, we present a brief Monte Carlo simulation to verify our mathematics. Finally, we discuss the limits of our model, and how experimentalists should best navigate calibration.

\section{\label{sec:level1}Derivations}

In this section, we attempt to mathematically model the statistical processes underlying the calibration scheme. We will largely confine our quantitative analysis to distributions' means and variances. The motivations for this are two-fold: for one, a calibration that requires only the measurement of a mean and standard deviation is more straightforward than one that requires a numerical fitting of distribution shapes. Secondly, it is not necessary that a distribution's probability density function be modelable by a closed-form function. Sticking to means and variances is therefore much more general.

In calculating the means and variances of dependent random variables, we make liberal use of the laws of total expectation and variance.

\subsection{\label{sec:level2}Classical Assumptions}
Let $\mathcal{N}$ be a random variable corresponding to the number of photons emitted in a given pulse of the light source. We expect for $\mathcal{N}$ to be distributed according to Poisson statistics, so we can characterize the mean and variance trivially:
\begin{equation}
    \label{eq:mu_poisson}
    \mu_{\mathcal{N}} = \lambda
\end{equation}

\begin{equation}
    \label{eq:var_poisson}
    \sigma^2_{\mathcal{N}} = \lambda
\end{equation}
Any given photon has an equal (and independent) probability $p$ of striking the detector when accounting for both attenuation along the fiber, geometric light collection efficiency, and reflection probabilities. Let $\mathcal{N}'$ model the number of photons that are actually absorbed by a detector. Recognizing this light collection efficiency as a binomial distribution conditional on $\mathcal{N}$, we find:

\begin{equation}
    \label{eq:mu_N_prime}
    \mu_{\mathcal{N'}} = \text{E}\left( \mathcal{N}'|\mathcal{N}\right)  = p\lambda
\end{equation}
\begin{equation}
\label{eq:var_N_prime}
\begin{split}
    \sigma^2_{\mathcal{N'}} =& \text{E}\left(\text{Var}\left( \mathcal{N}'|\mathcal{N}\right)\right) + \text{Var}\left(\text{E}\left( \mathcal{N}'|\mathcal{N}\right)\right)\\
    =& \text{E}\left(p(1-p)\mathcal{N} \right)+\text{Var}\left( p\mathcal{N}\right)\\
    =& p(1-p)\lambda +p^2\lambda = p\lambda
\end{split}
\end{equation}
The detector will be read out in physical units (e.g. current for a TES or phase for a KID) as opposed to number of photons. Denote $I_0$ the response in these units to a single photon being absorbed. Here, we assume that the detector is operating within the linear response regime and that signal size and the deposited energy are therefore directly proportional. The precision with which we can measure this response is limited by detector noise, as well as any sub-optimality in our reconstruction algorithm. We will assume that the distribution of our measured amplitude around the true amplitude due to these effects is independent of the amplitude's magnitude, and that it has variance $\sigma^2_i$. Consider now $\mathcal{I}$, corresponding to the distribution of measured amplitudes. Since our additional readout noise is uncorrelated to $\mathcal{N'}$, the distribution is simply scaled by $I_0$, and the variances sum:

\begin{equation}
    \label{eq:mu_I_1}
    \mu_{\mathcal{I}} = I_0p\lambda 
\end{equation}

\begin{equation}
\label{eq:var_I_1}
\begin{split}
    \sigma^2_{\mathcal{I}} =& I_0^2p\lambda + \sigma^2_i \\
\end{split}
\end{equation}

We can then combine equations \ref{eq:mu_I_1} and \ref{eq:var_I_1} in a more enlightening way:

\begin{equation}
\label{eq:final_naive}
\begin{split}
    \sigma^2_{\mathcal{I}} =& I_0 \mu_{\mathcal{I}} + \sigma^2_i \\
\end{split}
\end{equation}

As is apparent in eq. \ref{eq:final_naive}, the mean and variance of measured amplitudes are related linearly. Furthermore, the slope of this curve yields $I_0$, which, along with the photon energy, yields the detector's calibration coefficient, relating a pulse's amplitude to the amount of energy deposited in the substrate. The intercept corresponds to the squared amplitude resolution (in units of the readout amplitude). Combining the two yields the energy resolution of the detector.

The process of collecting $\left( \mu_\mathcal{I},\sigma^2_\mathcal{I}\right)$ is rather straightforward: the experimenter pulses the light source at some fixed intensity and collects data. Thereafter, they alter the light source's intensity and repeat the process. The user will have (for each intensity) a distribution of observed signal amplitudes, the mean and variance of which they can easily calculate. The relationship between mean and variance can easily be plotted and fit to a linear function, yielding measurements of both the calibration constant $I_0$ and the noise $\sigma_i$.

\subsection{A More Realistic Toy Model}
Key to the previous section's argument was the assumption that the detector resolution is uncorrelated to the number of photons that deposit energy in the detector. This is not realistic in practice; detector resolutions, for a variety of reason, generally increase with signal size. In this section, we'll develop a more realistic toy model of athermal phonon sensors that includes some of these noise sources, and investigate how they quantitatively affect our ability to perform the calibration described in section IIA. We will show that equation \ref{eq:final_naive} holds with a correction factor of $1 + \delta$ to $I_0$. We will conclude with a quantitative estimation of $\delta$.

To begin, we will confine ourselves to discussing events in which exactly one photon is absorbed by our detector. The absorption of this photon of energy $E_\gamma$ will first create electron-hole pairs and optical phonons in the substrate. Electron-hole pairs will recombine after some time, converting into optical phonons. These optical phonons have energies and momenta constrained by the substrate's phonon dispersion relations, and will promptly anharmonically decay into acoustic phonons defined by their own dispersion relations. These acoustic phonons are far out of thermal equilibrium with the substrate, and will be referred to as athermal phonons going forward. The athermal phonons will initially propagate diffusively, rapidly downconverting and scattering off of mass defects in the bulk substrate. After some time, the mean free path of the phonons grows larger than the length scale of the substrate, and they can be thought of as ballistic. During the transition from optical phonon to ballistic athermal phonon, we will make the simplifying assumption that a fraction $f_1$ of the total energy $E_\gamma$ is converted into measurable athermal phonons (i.e. athermal phonons with energy greater than than twice the superconducting bandgap $\Delta_{SC}$ of the sensor), with the remaining energy going towards subgap phonons that cannot be detected. 
Let $\mu_{\mathcal{E_\text{ph}}}$ and $\sigma^2_{\mathcal{E_\text{ph}}}$ denote the mean and variance in the energy of the detectable athermal phonons. The number of detectable athermal phonons produced by a single photon's absorption is given by $\mathcal{N}_\text{ph}$, with mean
\begin{equation}
    \label{eq:mu_n_ph}
    \mu_{\mathcal{N}_\text{ph}} = \frac{f_1E_{\gamma}}{\mu_{\mathcal{E}_\text{ph}}}
\end{equation}

and a variance that is sub-Poisson due to the correlated nature of the number of phonons created. We thus introduce the Fano Factor $F$ and the variance as  
\begin{equation}
    \label{eq:var_n_ph}
    \sigma^2_{\mathcal{N}_\text{ph}} = F\frac{f_1E_{\gamma}}{\mu_\mathcal{E_\text{ph}}} 
\end{equation}
where $F$ is some number less than one~\cite{Fanofactor}.

For now, we will assume that each of these phonons has a probability $q$ of reaching the sensor, breaking Cooper pairs, and depositing energy into the detector's sensing element. Note that $q$ is assumed to be independent of energy for detectable phonons. We can then study $\mathcal{N}^*_\text{ph}$, the distribution of the number of detected athermal phonons in a single-photon absorption:

\begin{equation}
    \label{eq:mu_n_ph_star}
    \mu_{\mathcal{N^*}_\text{ph}} = \text{E}\left(\mathcal{N^*}_\text{ph}|\mathcal{N}_\text{ph} \right) = q\frac{f_1E_\gamma}{\mu_{\mathcal{E}_\text{ph}}}
\end{equation}

\begin{equation}
    \label{var_n_ph_star}
    \begin{split}
    \sigma^2_{\mathcal{N^*}_\text{ph}} =& \text{E}\left(\text{Var}\left( \mathcal{N}^*_\text{ph}|\mathcal{N}_\text{ph} \right) \right) + \text{Var}\left(\text{E}\left( \mathcal{N}^*_\text{ph}|\mathcal{N}_\text{ph} \right) \right) \\
    =& \text{E}\left(q\left(1-q\right)\mathcal{N}_\text{ph} \right) + \text{Var}\left(q\mathcal{N}_\text{ph}\right) \\
    =& q\left(1-q\right)\frac{f_1E_\gamma}{\mu_{\mathcal{E}_\text{ph}}} + q^2F\frac{f_1E_\gamma}{\mu_{\mathcal{E}_\text{ph}}}
    \end{split}
\end{equation}
However, phonon collection may not be constant, varying as a function of position. Let $\mathcal{Q}$ be the distribution of such average efficiencies (as sampled by the photons) with mean and variance $\mu_\mathcal{Q}$ and $\sigma^2_\mathcal{Q}$ respectively. Under these conditions, we can calculate the mean and variance of $\mathcal{N}'_\text{ph}$, the number of collected athermal phonons with position dependence:

\begin{equation}
    \label{eq:mu_n_ph_prime}
    \mu_\mathcal{\mathcal{N'}_\text{ph}} = \mu_\mathcal{Q}\frac{f_1E_\gamma}{\mu_\mathcal{E_\text{ph}}}
\end{equation}

\begin{equation}
\label{eq:var_n_ph_prime}
\begin{aligned}
\sigma^2_{\mathcal{N'}_\text{ph}} 
&= \mathrm{E}\!\left[\mathrm{Var}\!\left(\mathcal{N'}_\text{ph}\mid \mathcal{Q}\right)\right]
 + \mathrm{Var}\!\left[\mathrm{E}\!\left(\mathcal{N'}_\text{ph}\mid \mathcal{Q}\right)\right] \\
&= \mathrm{E}\Bigg[
\mathcal{Q}(1-\mathcal{Q})
\frac{f_1 E_\gamma}{\mu_{\mathcal{E}_\text{ph}}}
+ F \mathcal{Q}^2
\frac{f_1 E_\gamma}{\mu_{\mathcal{E}_\text{ph}}}
\Bigg] \\
&\quad + \mathrm{Var}\!\left(
\mathcal{Q}
\frac{f_1 E_\gamma}{\mu_{\mathcal{E}_\text{ph}}}
\right) \\
&= \frac{f_1 E_\gamma}{\mu_{\mathcal{E}_\text{ph}}}
\Bigg[
\mu_\mathcal{Q}
- (1-F)\!\left(\sigma^2_\mathcal{Q} + \mu_\mathcal{Q}^2\right)
\Bigg] \\
&\quad + \sigma^2_\mathcal{Q}
\frac{f_1^2 E_\gamma^2}{\mu^2_{\mathcal{E}_\text{ph}}}
\end{aligned}
\end{equation}

We can now consider $\mathcal{N}^\dagger_\text{ph}$, the number of athermal phonons detected in a pulse of photons whose number is sampled randomly according to equations \ref{eq:mu_N_prime} and \ref{eq:var_N_prime}. 

\begin{equation}
    \label{eq:mu_n_ph_dag}
    \mu_{\mathcal{N}^\dagger_\text{ph}} = \text{E}\left(\mathcal{N}^\dagger_\text{ph}|\mathcal{N}' \right) = \mu_\mathcal{Q}p\lambda\frac{f_1E_\gamma}{\mu_{\mathcal{E}_\text{ph}}}
\end{equation}

\begin{equation}
\label{eq:var_n_ph_dag}
\begin{aligned}
\sigma^2_{\mathcal{N}^\dagger_\text{ph}} 
&= \mathrm{E}\!\left[\mathrm{Var}\!\left(\mathcal{N}^\dagger_\text{ph}\mid \mathcal{N}'\right)\right]
 + \mathrm{Var}\!\left[\mathrm{E}\!\left(\mathcal{N}^\dagger_\text{ph}\mid \mathcal{N}'\right)\right] \\
&= \mathrm{E}\Bigg[
\mathcal{N}' \Bigg(
\frac{f_1 E_\gamma}{\mu_{\mathcal{E}_\text{ph}}}
\Big[\mu_\mathcal{Q} 
- (1-F)\!\left(\sigma^2_\mathcal{Q} + \mu_\mathcal{Q}^2\right)\Big] \\
&\quad\quad
+ \sigma^2_\mathcal{Q}
\frac{f_1^2 E_\gamma^2}{\mu^2_{\mathcal{E}_\text{ph}}}
\Bigg)
\Bigg] \\
&\quad + \mathrm{Var}\!\left(
\mathcal{N}' \mu_\mathcal{Q}
\frac{f_1 E_\gamma}{\mu_{\mathcal{E}_\text{ph}}}
\right) \\
&= \frac{p \lambda f_1 E_\gamma}{\mu_{\mathcal{E}_\text{ph}}}
\Bigg[
\mu_\mathcal{Q} 
- (1-F)\!\left(\sigma^2_\mathcal{Q} + \mu^2_\mathcal{Q}\right) \\
&\quad\quad
+ \frac{f_1 E_\gamma}{\mu_{\mathcal{E}_\text{ph}}}
\left(\sigma^2_\mathcal{Q} + \mu^2_\mathcal{Q}\right)
\Bigg]
\end{aligned}
\end{equation}

Finally, each of these detected phonons has a variable amount of energy described by $\mathcal{E_\text{ph}}$, with a fraction $f_2$ of that energy eventually thermalizing in the sensing element. We remark that $f_1f_2\mu_\mathcal{Q}$ satisfies our earlier definition of energy collection efficiency (see appendix A for a deeper discussion of the physical origins of these three parameters). Let $\mathcal{E_\text{tot}}$ describe the distribution of the total energy deposited in the sensing element by phonons:

\begin{equation}
    \label{eq:mu_e_tot}
    \mu_{\mathcal{E}_\text{tot}} = \text{E}\left(\mathcal{E}_\text{tot}|\mathcal{E}_\text{ph} \right) = \mu_\mathcal{Q}f_1f_2p \lambda E_\gamma
\end{equation}

\begin{equation}
\label{eq:var_e_tot}
\begin{aligned}
\sigma^2_{\mathcal{E}_\text{tot}} 
&= \mathrm{E}\!\left[\mathrm{Var}\!\left(\mathcal{E}_\text{tot}\mid\mathcal{E}_\text{ph}\right)\right]
 + \mathrm{Var}\!\left[\mathrm{E}\!\left(\mathcal{E}_\text{tot}\mid\mathcal{E}_\text{ph}\right)\right] \\
&= f_2^2 \mu^2_{\mathcal{E}_\text{ph}} \sigma^2_{\mathcal{N}^\dagger_\text{ph}}
 + f_2^2 \sigma^2_{\mathcal{E}_\text{ph}} \mu_{\mathcal{N}^\dagger_\text{ph}} \\
&= \mu_{\mathcal{E}_\text{ph}}\, p \lambda f_1 f_2^2 E_\gamma \Bigg[
\mu_\mathcal{Q} 
- (1-F)\!\left(\sigma^2_\mathcal{Q} + \mu^2_\mathcal{Q}\right) \\
&\quad\quad
+ \frac{f_1 E_\gamma}{\mu_{\mathcal{E}_\text{ph}}}
\left(\sigma^2_\mathcal{Q} + \mu^2_\mathcal{Q}\right)
\Bigg] \\
&\quad + \sigma^2_{\mathcal{E}_\text{ph}}\, \mu_\mathcal{Q}\, p \lambda
\frac{f_1 f_2^2 E_\gamma}{\mu_{\mathcal{E}_\text{ph}}}
\end{aligned}
\end{equation}
$\mathcal{E}_\text{tot}$ will be proportional to the total signal size observed in the detector. However, the detector will be read out in units of some amplitude rather than energy. Let $I_0$ once again be the mean response to a single photon, $\sigma_i^2$ the extra variance due to baseline fluctuations and sub-optimal reconstruction, and $\mathcal{I}$ the distribution of reconstructed amplitudes:

\begin{equation}
    \label{eq:mu_I_2}
    \mu_{\mathcal{I}} = I_0p\lambda
\end{equation}

\begin{equation}
\label{eq:var_I_2}
\begin{aligned}
\sigma^2_{\mathcal{I}} &= I^2_0 p \lambda \Bigg[
\frac{\mu_{\mathcal{E}_{\text{ph}}}}{f_1 E_\gamma}
\left(
\frac{1}{\mu_\mathcal{Q}}
- (1 - F)\left(1 + \frac{\sigma^2_\mathcal{Q}}{\mu^2_\mathcal{Q}}\right)
\right) \\
&\quad + \left(1 + \frac{\sigma^2_\mathcal{Q}}{\mu^2_\mathcal{Q}}\right) + \frac{1}{\mu_\mathcal{Q}}
\frac{\sigma^2_{\mathcal{E}_{\text{ph}}}}{f_1 E_\gamma \mu_{\mathcal{E}_{\text{ph}}}}
\Bigg]
+ \sigma_i^2
\end{aligned}
\end{equation}
Combining the two:

\begin{equation}
    \label{eq:final_correct}
    \sigma^2_\mathcal{I} = I_0\left(1+\delta\right)\mu_\mathcal{I} + \sigma_i^2
\end{equation}
where
\begin{equation}
\label{eq:delta}
\begin{aligned}
\delta 
&= \frac{\mu_{\mathcal{E}_\text{ph}}}{f_1 E_\gamma}
\Bigg[
\frac{1}{\mu_\mathcal{Q}}
- (1-F)\!\left(1 + \frac{\sigma^2_\mathcal{Q}}{\mu^2_\mathcal{Q}}\right)
\Bigg] \\
&\quad + \frac{\sigma^2_\mathcal{Q}}{\mu^2_\mathcal{Q}}
+ \frac{1}{\mu_\mathcal{Q}}
\frac{\sigma^2_{\mathcal{E}_\text{ph}}}{f_1 E_\gamma \mu_{\mathcal{E}_\text{ph}}}
\end{aligned}
\end{equation}

Although the variance and mean of this distribution are still linearly related, the slope of the relationship no longer faithfully corresponds to $I_0$, with deviations of order $\delta$. Naively assuming that the slope of this relation corresponds to $I_0$ (as in section IIA) will therefore yield an incorrect calibration, underestimating both the amount of energy in any given event and the baseline energy resolution of the sensor. Thus, if the effects of $\delta$ are large, previous measurements using this technique are incorrect. 



\section{\label{sec:level1}Simulations and Estimations }

\subsection{Bounds on Contributions to $\delta$}

In this section, we will take a comprehensive approach to estimating the impact of different detector parameters on $\delta$. Since there are many parameters we must consider ($f_1$, $F$, $\mu_\mathcal{Q}$, $\sigma^2_\mathcal{Q}$, $\mu_{\mathcal{E}_\text{ph}}$,$\sigma^2_{\mathcal{E}_\text{ph}}$), we will attempt to isolate the impact of individual parameters on $\delta$. In practice, we will calculate the value of $\delta$ while varying two of the above parameters simultaneously. The remaining parameters, left constant, are selected to minimize their contribution to $\delta$ (thus ensuring that $\delta$ is estimated conservatively) without violating physical law. The scanned parameters are allowed to vary over all physical parameter space. For this exercise, we consider a detector substrate made of silicon, with the relevant minimum detectable energy given by the superconducting band gap of aluminum, which represents the fin material in the common QET sensor layout~\cite{irwinQuasiparticleTrapAssisted1995}. Finally, the photon energy is simulated to be 3 eV.

\subsubsection{$\delta$ vs Energy Collection Efficiency}

\begin{figure}[b]
\includegraphics[width=\columnwidth]{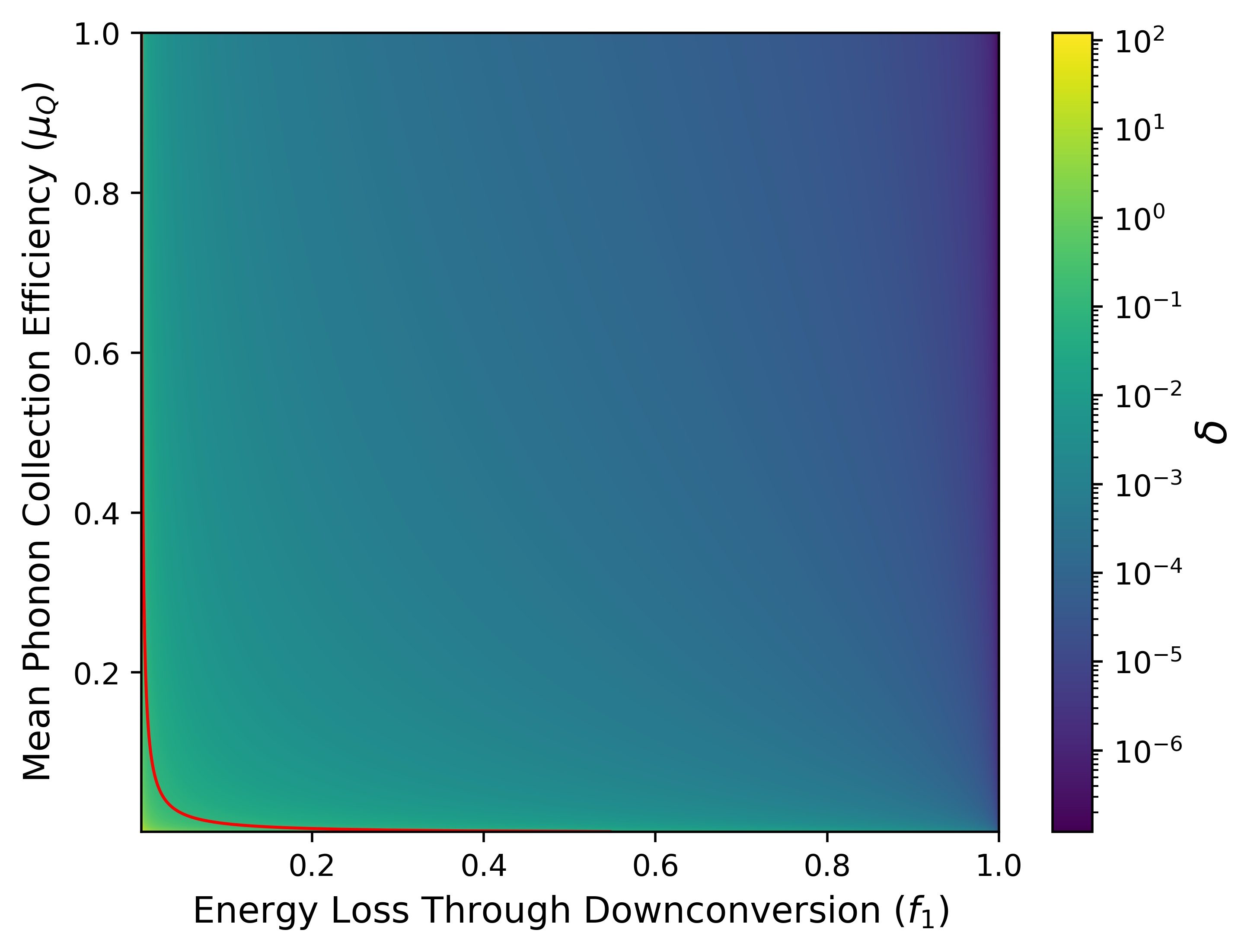}
\caption{\label{fig:delta_vs_muQ_f1} $\delta$ for various values of $\mu_Q$ and $f_1$. See text for description.}
\end{figure}

We begin with a scan in $\mu_\mathcal{Q}$ and $f_1$, the two parameters in $\delta$ that describe the energy collection efficiency of our sensor. Scans are performed over $\mu_\mathcal{Q}\in\left[.001,1\right]$ and $f_1\in\left[0,1\right]$. We conservatively fix the other parameters at $\sigma^2_\mathcal{Q}~=0$, $\mu_{\mathcal{E}_\text{ph}}=2\delta_{\text{Al}}$, $\sigma^2_{\mathcal{E}_\text{ph}}=0$, and $F=0$. Figure \ref{fig:delta_vs_muQ_f1} shows the corresponding scan, with a contour corresponding to $\delta~=.1$ superimposed in red. The vast majority of parameter space has $\delta < .1$, indicating that poor energy collection efficiency alone may not preclude  $\delta \approx0$. (It should however be noted that our minimal estimation of $\mu_{\mathcal{E}_\text{ph}}$ certainly underestimated $\delta$).

\subsubsection{$\delta$ vs Position Dependence}

\begin{figure}[b]
\includegraphics[width=\columnwidth]{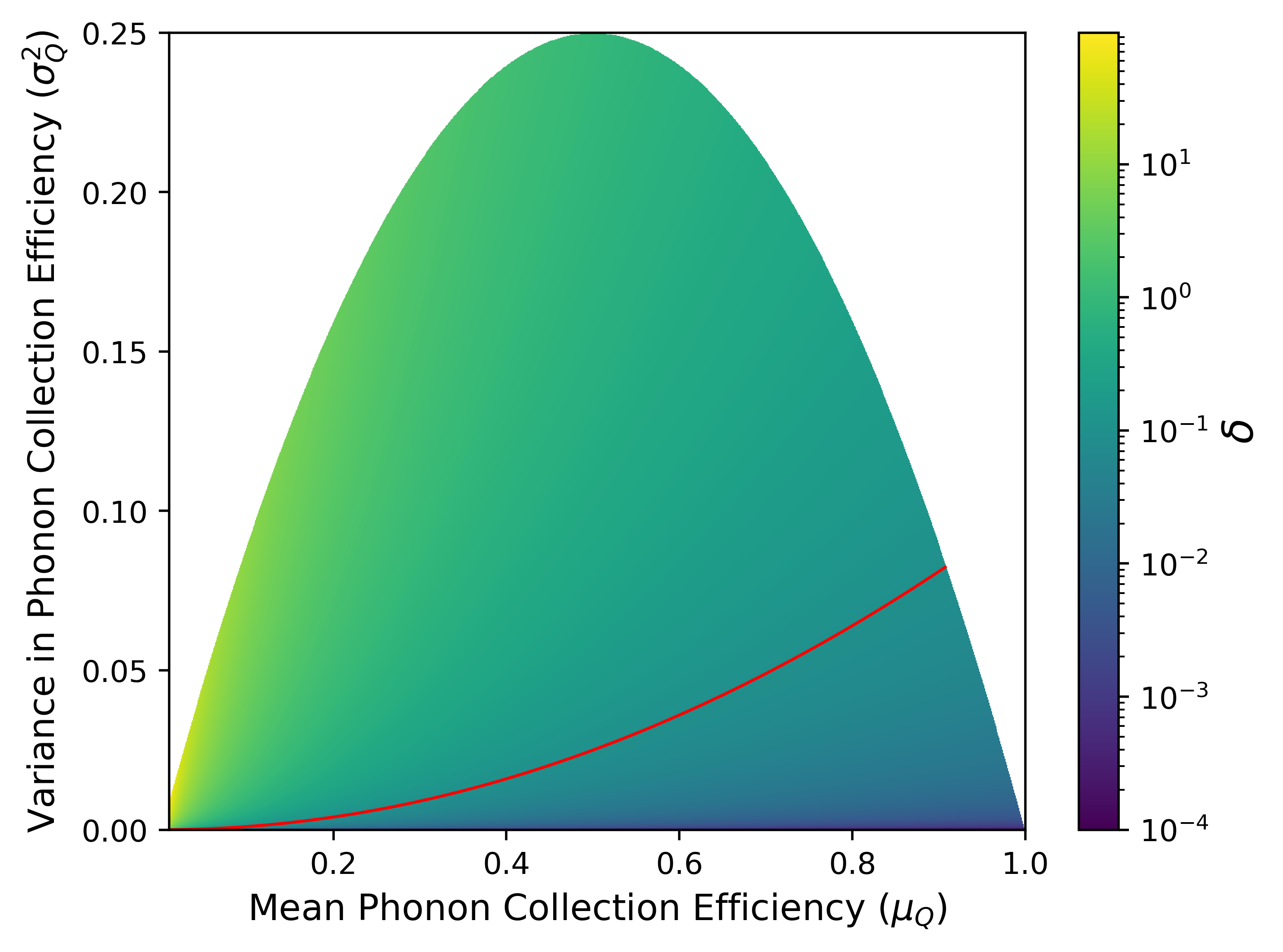}
\caption{\label{fig:delta_vs_muQ_sigQ} $\delta$ for various values of $\mu_Q$ and $\sigma^2_\mathcal{Q}$. See text for description.}
\end{figure}

Next we scan $\mu_\mathcal{Q}$ and $\sigma^2_\mathcal{Q}$, which together parameterize the relative magnitude of position dependence. Scans are performed over $\mu_\mathcal{Q}\in\left[.001,1\right]$ and $\sigma^2_\mathcal{Q}\in\left[0,\mu_\mathcal{Q}(1-\mu_\mathcal{Q})\right]$. Since $\mathcal{Q}$ is itself bound on $[0,1]$, its variance at fixed mean is saturated in the limit of a Bernoulli distribution. We conservatively fix the other parameters at $f_1=1$, $\mu_{\mathcal{E}_\text{ph}}=2\Delta_{\text{Al}}$, $\sigma^2_{\mathcal{E}_\text{ph}}=0$, and $F=0$. Figure \ref{fig:delta_vs_muQ_sigQ} shows the scan under these conditions. The vast majority of available parameter space has $\delta>.1$; it is only for relatively high phonon collection efficiency and low position dependence that $\delta\approx0$ is maintained. We therefore note that position dependence (especially in conjunction with a low average phonon collection efficiency) is a large barrier to achieving $\delta \approx 0$.

\subsubsection{$\delta$ vs Phonon Energy}

\begin{figure}[b]
\includegraphics[width=\columnwidth]{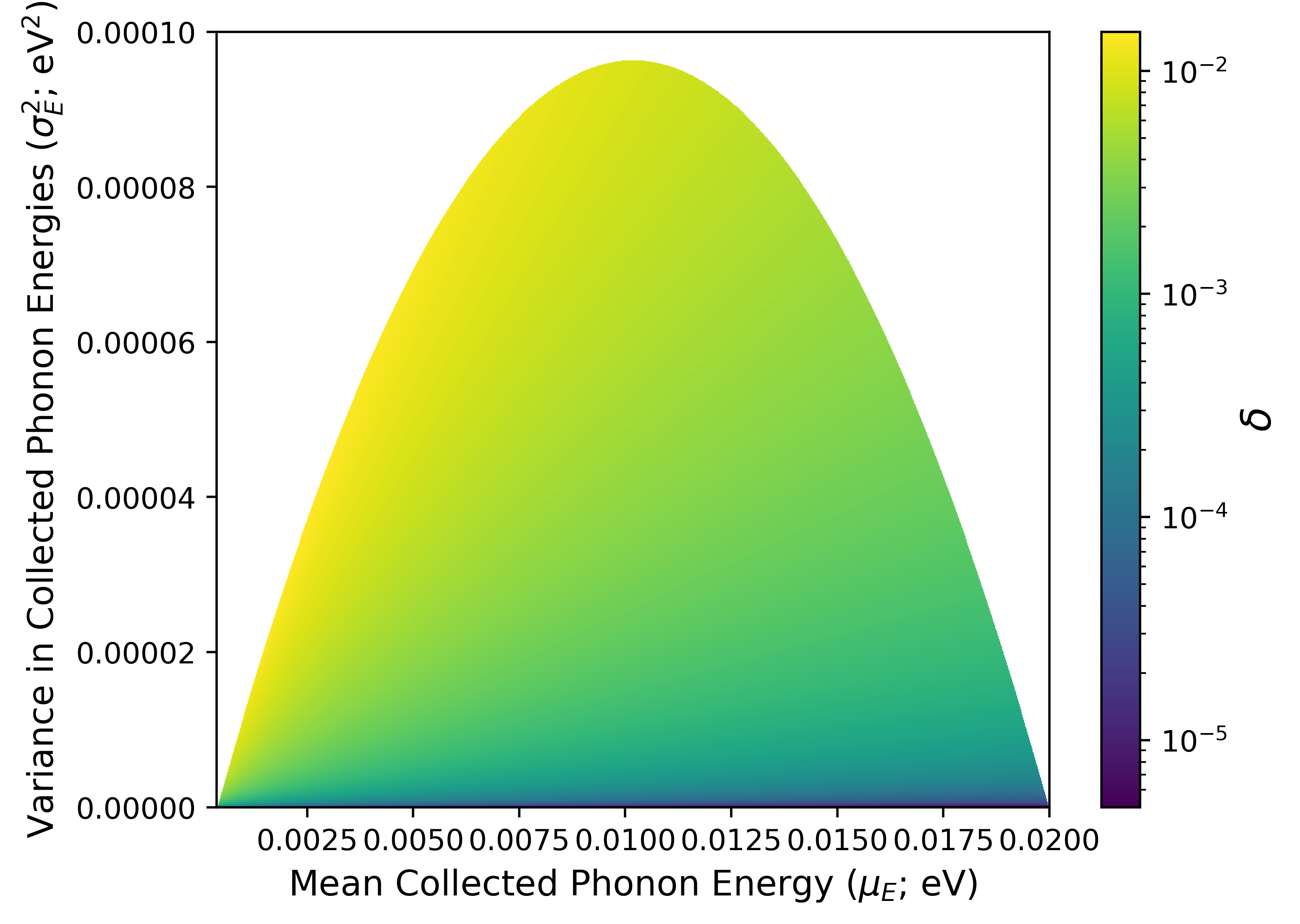}
\caption{\label{fig:delta_vs:muph_sigph} $\delta$ for various values of $\mu_{\mathcal{E}_\text{ph}}$ and $\sigma^2_{\mathcal{E}_\text{ph}}$. See text for description.}
\end{figure}

Next we scan in $\mu_{\mathcal{E}_\text{ph}}$ and $\sigma^2_{\mathcal{E}_\text{ph}}$, the parameters describing the energy of athermal phonons that go on to break Cooper pairs in our sensing element. Scans are performed over $\mu_{\mathcal{E}_\text{ph}} \in \left[2\Delta_\text{Al},4E_\text{bal} \right]$ and $\sigma^2_{\mathcal{E}_\text{ph}} \in \left[0,\left(\mu_{\mathcal{E}_\text{ph}}-2\Delta_\text{Al}\right)\left(4E_\text{bal} -\mu_{\mathcal{E}_\text{ph}}\right) \right]$. The former is practically bound between the energy required to break Cooper pairs in the aluminum sensing element and a few times $E_\text{bal}$, the upper bound for ballistic phonon propagation in silicon, around 5 meV \cite{SiBallCutoff}. The latter is bound from above by the maximum achievable variance for a distribution bound on $\left[2\Delta_\text{Al},4E_\text{bal} \right]$ with fixed mean $\mu_{\mathcal{E}_\text{ph}}$. The other parameters are conservatively set to $\mu_\mathcal{Q}=1$, $\sigma^2_\mathcal{Q}=0$, $f_1=1$, and $F=0$. The result of this scan is found in Fig. \ref{fig:delta_vs:muph_sigph}. For all available parameter space, $\delta < .1$, indicating that excess variance due to athermal phonon energy variance is unlikely to contribute to large values of $\delta$.

\subsubsection{$\delta$ vs Fano Fluctuations}

\begin{figure}[b]
\includegraphics[width=\columnwidth]{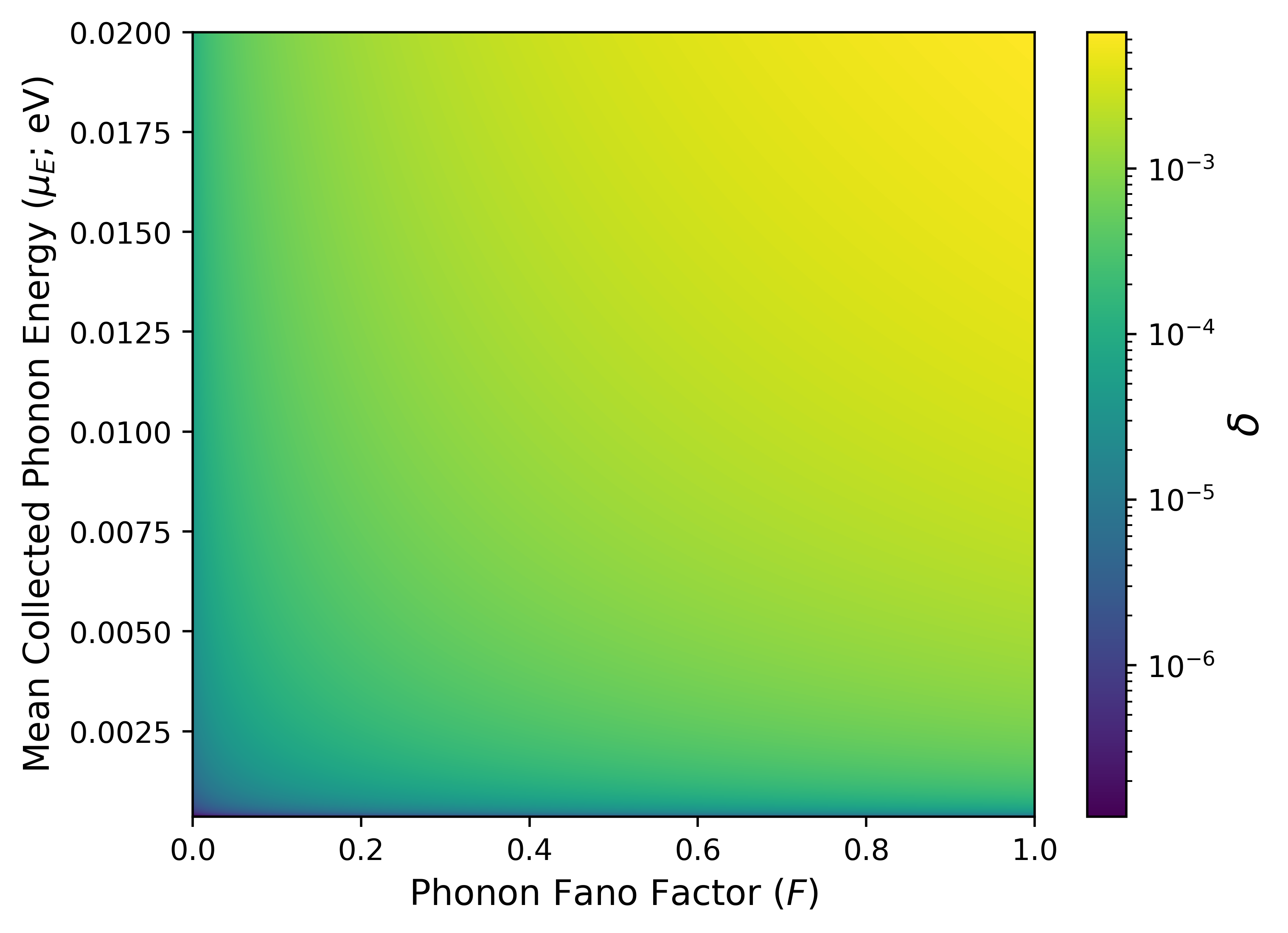}
\caption{\label{fig:delta_vs_Eph_F} $\delta$ for various values of $\mu_{\mathcal{E}_\text{ph}}$ and $F$. See text for description.}
\end{figure}

Finally we scan in $F$ and $\mu_{\mathcal{E}_\text{ph}}$, which together characterize the signal variance due to Fano fluctuations in the number of generated athermal phonons. Scans are performed over $F \in \left[0,1 \right]$ and $\mu_{\mathcal{E}_\text{ph}} \in \left[2\Delta_\text{Al},4E_\text{bal} \right]$. The other parameters are conservatively set to $f_1=1$, $\mu_\mathcal{Q}=1$, $\sigma^2_\mathcal{Q} =0$, and $\sigma^2_{\mathcal{E}_\text{ph}}=0$. Fig. \ref{fig:delta_vs_Eph_F} shows the corresponding scan. Since $\delta<.1$ for all available parameter space, it is unlikely that Fano fluctuations in the number of athermal phonons generated alone will contribute much to $\delta$.

\subsection{Example Simulation}

In this section, we will use simple Monte Carlo simulations to illustrate the implications of the above mathematics on calibration in practice. We simulate photon and phonon emission and detection, detector noise, and position dependence by drawing from appropriate random variables as described below.

We begin by modeling the light source. The photons are modeled as perfectly monoenergetic at 3 eV, with the number of detector photons in each pulse of light ($p\lambda$) generated according to a Poisson random variable. Variance in phonon collection efficiency is modeled as a uniform distribution with multiple means ($\mu_\mathcal{Q}$) and variances ($\sigma^2_\mathcal{Q}$) tested. We simulate four different calibration scenarios. Scenario 1 exhibits a phonon collection efficiency of ($\mu_\mathcal{Q}=.8$), a relatively high efficiency, and no position dependence ($\sigma^2_\mathcal{Q}=0$). Scenario 2 exhibits the same collection efficiency ($\mu_\mathcal{Q}=.8$) with substantial position dependence ($\sigma^2_\mathcal{Q}=.0133$), such that $\mathcal{Q}$ corresponds to a uniform distribution between $.6$ and $1$. Scenario 3 exhibits a low phonon collection efficiency ($\mu_\mathcal{Q}=.05$) with no position dependence ($\sigma^2_\mathcal{Q}=0$). Scenario 4 exhibits the same low phonon collection efficiency ($\mu_\mathcal{Q}=.05$) with a substantial, albeit not maximal position dependence ($\sigma^2_\mathcal{Q}=0.000833$), making $\mathcal{Q}$ correspond to a uniform distribution between $0$ and $.1$. For each scenario, $1000$ pulses were simulated at light source intensities corresponding to $p\lambda=[100, 200,...,900]$ mean generated photons. Phonon energies are generated according to a uniform distribution bound from above by silicon's ballistic cutoff energy $.005$ eV \cite{SiBallCutoff} and from below by the superconducting band gap for aluminum. The Fano factor for phonon generation is assumed to be $F=1$. Other losses are modeled as $f_1=f_2=0.8$. Finally, the detector is assumed to have calibration constant $I_0=1 \text{ nA/eV}$ and current noise of $\sigma_i=1 \text{ nA}$.

\begin{figure}[b]
\includegraphics[width=\columnwidth]{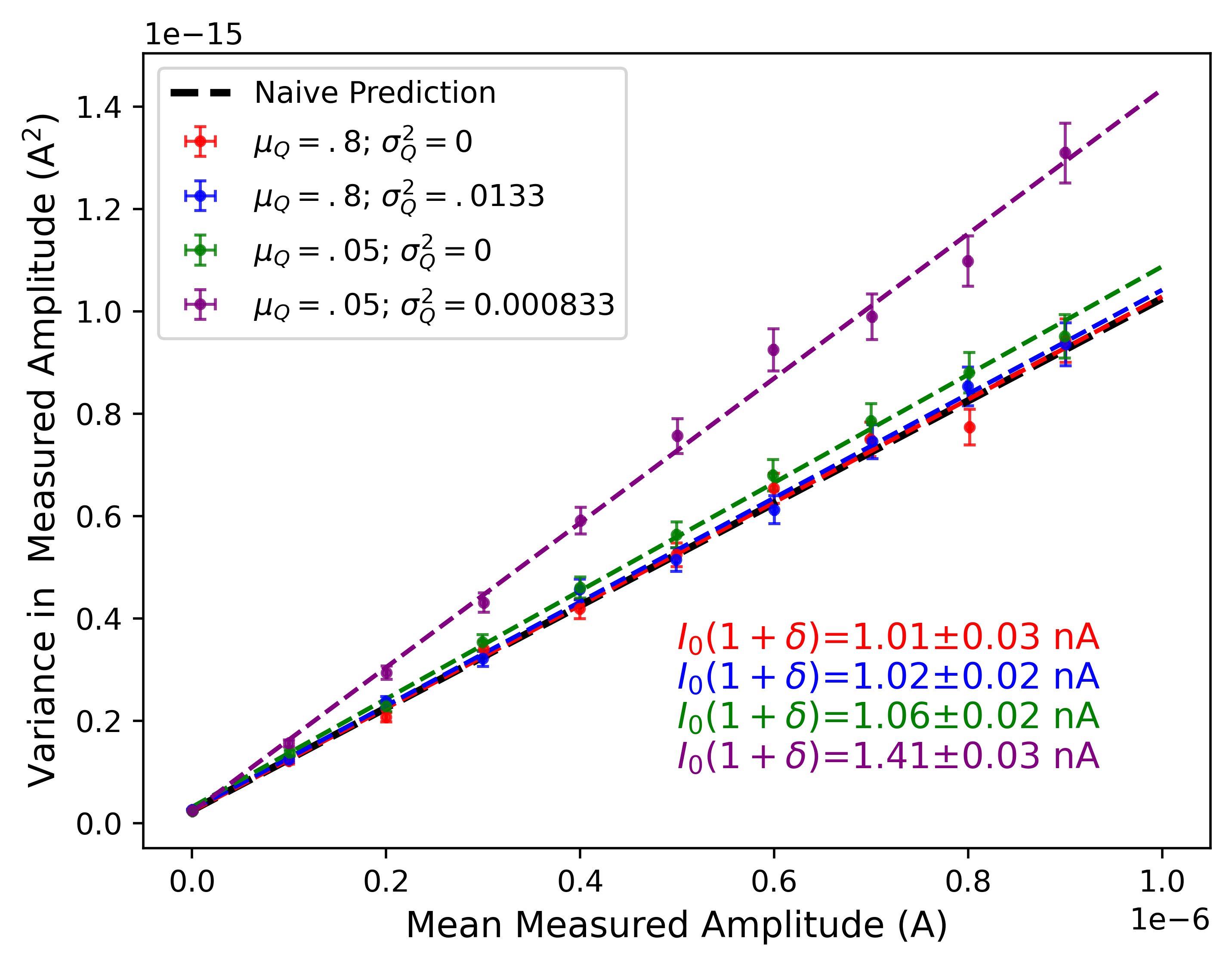}
\caption{\label{fig:example_calib} Plot of variances vs means of amplitudes for the four simulated detector amplitudes, along with best-fit lines and the expected fit under the naive assumptions laid out in section IIA.}
\end{figure}

Figure \ref{fig:example_calib} displays the results of this simulation. As can be seen, despite all four detectors having the same nominal calibration constant $I_0$ and readout noise $\sigma_i$, the measured $I_0\left(1+\delta\right)$ differ. The greatest departure from the naive expectation (i.e. largest $\delta$) is observed for the detector with low-phonon collection efficiency and significant position dependence. The observed slopes also agree well with the values predicted using equation \ref{eq:delta}.

\section{\label{sec:level1}Discussion}

While the mathematics in sections II/III are simple, we emphasize that their application to the calibration of cryogenic particle detectors has largely been ignored in literature \cite{Cardani_2018,Calder_Final,TemplesKID,BullKID_orig,Nucleus2Channel,LANTERN}. Most authors follow the procedure laid out in section IIA, implicitly assuming that $\delta=0$ without further consideration. We will discuss when this is valid, and the bounds of our model.

In this work we have made simplifications to the processes governing the decay of phonons and the various loss mechanisms that lead to imperfect energy collection. This was done to simplify the mathematical model presented above while still describing all of the relevant physics (in Appendix A we describe how our models parameterization of energy loss corresponds to more detailed description energy loss inside the detector). Where we've made approximations, we've attempted to ensure that they underestimate $\delta$ in order to assure the result is maximally conservative. For example, we've modeled fractional loss of an athermal phonon's energy in the substrate and sensing element as constant ratios, $f_1$ and $f_2$ respectively. In reality, these numbers are not constant for all phonons, and might be better modeled as being drawn from  distributions with nonzero variance. However, by ignoring these variances, we've underestimated the eventual variance of $\mathcal{I}$, and therefore underestimated $\delta$. 

One exception to this philosophy has been in our treatment of covariance between distributions. Barring explicit dependencies, we've treated all distributions as being sampled independently. If two distributions are instead anti-correlated, then we risk overestimating $\delta$. $\mathcal{E}_\text{ph}$ and $\mathcal{N}_\text{ph}$ are obvious candidates for distributions with anticorrelations by nature of conservation of energy. This effect will be mitigated by the fact that $\mathcal{E_\text{ph}}$ is the energy of athermal phonon after all downconversion processes and not the original phonon's energy. The loss mechanisms implicit in downconversion will muddle the otherwise perfect correlation between phonon energy and number. In addition, we showed in section III that effects from athermal phonon energy variance and Fano fluctuations are in most circumstances subdominant to those from energy collection efficiency and position dependence. We can therefore take the limit of $\sigma^2_{\mathcal{E}_\text{ph}}=0$ and $F=0$ and still have a conservative estimate of $\delta$. In this limit, our result regarding contributions from energy collection efficiency and position dependence remain maximally conservative, and still potentially large enough to merit concern

Based on the results of section III, we predict that state-of-the art superconducting sensors with high energy collection efficiency (e.g. QETs) are likely safe to calibrate using the methods of section IIA, and that any systematic error stemming from $\delta\neq0$ are likely to be of order $10\%$ or less. Ironically, it is with these detectors that achieving sub-single-photon resolution is possible, allowing a calibration as described in Fig. \ref{fig:subphot_res}. 

In contrast, more emergent technologies, such as KIDs and qubit-based detectors have lower phonon collection efficiencies due to large dead metal areas absorbing athermal phonons~\cite{TemplesKID,celi2026measuringquasiparticledynamicsparticle} and as yet unsuccessful implementation of quasiparticle trapping schemes \cite{KIPM_update}. These not only cause poor phonon collection, but also increase the amount of position dependence in these detectors. As an example, a recoil that occurs in the substrate directly below a sensing element will have a greater phonon collection efficiency than one below a passive metal film (e.g. a grounding plane or resonator feedline)~\cite{Wen2025}. Photon-based calibration of these detectors cannot be viewed as accurate without considerable effort towards understanding the effects of $\delta$. Moreover, quoted energy resolutions of such sensors measured by photon calibration should be interpreted strictly as optimistic lower bounds to the true resolutions. Finally, the stated sensor thresholds (often set as some multiple of the energy resolution to define the trigger significance) are proportionally optimistic. Ultimately, any photon-based calibration for such a sensor should be accompanied by an x-ray calibration, not only to give confidence to the detector's characterization, but also to give credibility to any measurements (e.g.\ dark matter limits) being set with the detector.

Detailed Monte Carlo simulation of phonon dynamics remains a possible path to estimating the value of $\delta$. Tools such as G4CMP \cite{G4CMP,G4CMP_novel} have the potential to bound or approximate the contributions to $\delta$ stemming from phonon dynamics. In fact, work has already been published with direct relevance to estimating energy collection efficiency in KID-based detector \cite{KIPM_update}, which could be applied to constraining/estimating $\delta$. However, this work overestimated the energy collection efficiency by an order of magnitude, indicating that more development is necessary before $\delta$ can be predicted \textit{a priori} in simulation with sufficient precision.

Another path toward estimating $\delta$ exists in steerable beam-like cryogenic photon sources recently developed explicitly for calibrating superconducting calorimeters \cite{Stifter_MEMS,Kurinsky_MEMS}. By using such a source and calibrating at discrete points along the a detector, the effect of $\sigma^2_\mathcal{Q}$ can effectively be removed (assuming that the phonon collection efficiency is roughly constant on length scales corresponding to the source's spot size.) Moreover, a raster scan over the detector's entire area will actually enable a measurement of $\sigma^2_\mathcal{Q}/\mu^2_\mathcal{Q}$, the most troublesome term in eq. \ref{eq:delta}.

We conclude our discussions by noting that potential experimental evidence for $\delta\neq0$ has been published by the BullKID collaboration in the form of a few \% discrepancy between calibrations produced by 3 eV optical phonons and 60 keV gamma rays \cite{BullKID_calibration}. While they cite Poisson fluctuations in the number of generated phonons as the likely culprit, section III of this work indicates that the more-likely cause is position dependence within each silicon voxel.

\section{Conclusion}
In this work, we have examined the statistical foundations of photon-based calibration techniques for cryogenic calorimetric detectors and demonstrated that the commonly assumed calibration scheme can be systematically biased when detector non-idealities are taken into account. By considering fluctuations in phonon production, transport, and collection, we have derived a correction factor, $\delta$, that estimates deviations from the ideal calibration slope. We show that $\delta$ can be non-negligible across broad regions of reasonable parameter space, especially for detectors with low phonon collection efficiency and significant position dependence. These results imply that naive application of Poisson-based calibration methods may lead to systematic misestimation of both the detector energy scale and resolution, especially so for emerging detector technologies like KIDs and qubit-based sensors.

Looking forward, incorporating detailed phonon transport modeling and experimental measurements of collection efficiency and position dependence will be critical steps toward placing further bounds on the effect of $\delta$. Understanding and reducing these systematics will enable the reliable calibration of next-generation low-energy detectors.

\begin{acknowledgments}
We thank Xinran Li, Dan McKinsey, and Peter Sorensen for their input on this work. We thank Ryan Linehan for useful discussions regarding the phonon simulation package G4CMP. This work was supported by the U.S. Department of Energy (DOE) Office of Science, Office of High Energy Physics under contract DE-AC02-05CH11231. This material is based upon work supported by the Department of Energy National Nuclear Security Administration through the Nuclear Science and Security Consortium under Award Number DE-NA0003996. This report was prepared as an account of work sponsored by an agency of the United States Government. Neither the United States Government nor any agency thereof, nor any of their employees, makes any warranty, express or implied, or assumes any legal liability or responsibility for the accuracy, completeness, or usefulness of any information, apparatus, product, or process disclosed, or represents that its use would not infringe privately owned rights. Reference herein to any specific commercial product, process, or service by trade name, trademark, manufacturer, or otherwise does not necessarily constitute or imply its endorsement, recommendation, or favoring by the United States Government or any agency thereof. The views and opinions of authors expressed herein do not necessarily state or reflect those of the United States Government or any agency thereof.
\end{acknowledgments}

\appendix

\section{Simulation of Energy Loss in the Detector}
\label{appendixA}
Section IIB sets forth a relatively simple model of losses throughout the detection chain. In this appendix, we'll motivate the qualitative adherence of this model to reality.

Phonon generation begins with particle interaction in the substrate material, generating both electron-hole pairs and optical phonons, with the former population themselves decaying to optical phonons. Optical phonons will promptly downconvert into longer-lived acoustic phonons \cite{srivastava1990phonons}. The acoustic phonons can decay and scatter on crystal defects and impurities (both chemical and isotopic) in the bulk substrate, as well as anharmonically downconvert into multiple lower energy acoustic phonons spontaneously \cite{srivastava1990phonons}.

Not all of the initial energy will arrive in the sensing element/collection fins. The main loss mechanism is downconversion to phonons with insufficient energy to break Cooper pairs, ensuring that the resultant phonon is bound to the substrate. Loss through downconversion can occur in two forms: in the first case, a high-energy acoustic phonon downconverts into another high-energy phonon and a sub-gap phonon. This manifests as a slight attenuation in the original phonon's energy. In the second case, a near-gap phonon downconverts into two sub-gap phonons, precluding detection of even a fraction of the original phonon's energy. In the text of the paper, we quantify the former loss using $f_1$, while the latter is baked into the phonon collection efficiency $q$ (and later $\mu_\mathcal{Q}$). Pair-breaking in passive metal films, causing total loss of an athermal phonon's energy, is also considered in $\mathcal{Q}$.

Each athermal phonon with sufficient energy has some probability of reaching the superconducting sensing element. In the case of a QET-like architecture, this volume consists of the large-area superconducting fins, where an incident athermal phonon can break a Cooper pair, generating a Bogoliubov quasiparticle \cite{irwinQuasiparticleTrapAssisted1995}. This quasiparticle can then go on to break more Cooper pairs (if it has sufficient energy), while simultaneously radiating away lower-energy phonons. This results in one or more Bogoliubov quasiparticles with energy near the gap edge, and a bath of low energy athermal phonons in the collection fins. The former propagate diffusively with a bias against the bandgap gradient (i.e. preferentially toward the TES), while the latter propagate isotropically. As the quasiparticles reach lower gap regions, they can shed more acoustic phonons, trapping them. Upon recombination, a phonon is emitted which can either re-break Cooper pairs, propagate away, or downconvert and eventually thermalize.

Loss in the sensing elements occurs in two forms: in the first, the initial Boguliubov quasiparticle can recombine in the collection fins, preventing energy from being funneled into the TES. This manifests as none (or very little) of the incident phonon's energy reaching the TES. The second case results in only a fraction of the incident phonon's energy reaching the TES. This manifests when a fraction of the the radiated athermal phonons do not thermalize in the TES, or when a secondary Boguliubov quasiparticle recombines in collection fins. The case in which all of the acoustic phonon's energy is lost is considered in $q$ and $\mu_\mathcal{Q}$, while the case in which only a fraction is lost is quantified using $f_2$.

\section{Non-Poisson Photon Emission}

In this section, we present a simple addendum to the argument reproduced in section II, extending its use to the case in which the pulsed light source's photon emission statistics are not Poisson. 

Let $\mathcal{N}$ once again be the number of photons emitted in a given pulse. Since $\mathcal{N}$ is no longer faithfully modeled by a Poisson distribution, we will write generally:

\begin{equation}
    \label{eq:mu_N_nonpoisson}
    \mu_\mathcal{N}=\lambda
\end{equation}

\begin{equation}
    \label{eq:var_N_nonpoisson}
    \sigma^2_\mathcal{N}=\sigma_N^2
\end{equation}

After attenuation, with each generated photon having probability $p$ of being absorbed, we have a distribution $\mathcal{N}'$ of absorbed photons in a pulse:

\begin{equation}
\label{eq:mu_N_prime_nonpoisson}
    \mu_{\mathcal{N'}}=\text{E}\left(\mathcal{N'}|\mathcal{N}\right)=p\lambda
\end{equation}

\begin{equation}
\label{eq:var_N_prime_nonpoisson}
\begin{split}
    \sigma^2_{\mathcal{N'}} =& \text{E}\left(\text{Var}\left( \mathcal{N}'|\mathcal{N}\right)\right) + \text{Var}\left(\text{E}\left( \mathcal{N}'|\mathcal{N}\right)\right)\\
    =& \text{E}\left(p(1-p)\mathcal{N} \right)+\text{Var}\left( p\mathcal{N}\right)\\
    =& p(1-p)\lambda +p^2\sigma^2_N = p\lambda +p^2\left(\sigma^2_N-\lambda\right)
\end{split}
\end{equation}

Maintaining our assumptions about responsitivity and noise, we can calculate the mean and variance of $\mathcal{I}$, the distribution of measured pulse amplitudes:

\begin{equation}
    \label{eq:mu_I_nonpoisson}
    \mu_\mathcal{I} = I_0p\lambda
\end{equation}

\begin{equation}
    \label{eq:var_I_nonpoisson}
    \sigma^2_\mathcal{I} = I_0^2p\lambda + I_0^2p^2\left(\sigma^2_N-\lambda\right)+\sigma_i^2
\end{equation}

While we'd like to rewrite equation \ref{eq:var_I_nonpoisson} using equation \ref{eq:mu_I_nonpoisson}, the extra quadratic term precludes this: we do not how $\sigma^2_N-\lambda$ varies with N. We will consider two possible scenarios.

In the first case, $\mathcal{N}$ is "quasi-Poisson", meaning that its mean and variance are  proportional for all $\lambda$. Writing the constant of proportionality as $1+\epsilon$, we can rewrite \ref{eq:var_I_nonpoisson} as desired:

\begin{equation*}
    \label{eq:final_linear_non_poisson}
    \sigma^2_\mathcal{I} = I_0\left(1+p\epsilon\right)\mu_\mathcal{I}+\sigma_i^2
\end{equation*}

While $\sigma^2_\mathcal{I}$ and $\mu_\mathcal{I}$ are still linearly related, the slope no longer corresponds to the calibration constant.

In the second case, $\mathcal{N}$ is not quasi-Poisson, meaning that $\sigma^2_N-\lambda$ is not simply linear in $\lambda$. Nor then will $\sigma^2_\mathcal{I}$ and $\mu_\mathcal{I}$ be linearly related, a fact that should be immediately obvious to the experimenter.

To reiterate our dichotomy: in the former case, the calibration is biased in a way that the user has no sensitivity to, while in the latter, the user should immediately know that the calibration is incorrect. While the latter is certainly preferable, in either case the calibration produced is invalid.

Comparing equations \ref{eq:var_I_nonpoisson} and \ref{eq:var_I_1}, we note that the additional, problematic term in \ref{eq:var_I_nonpoisson} is the only one quadratic in $p$. While we are not sensitive to the true value of $p$ (in the same way that we are not sensitive to the true number of photons being produced inside our light source,) we can split $p$ into an effective attenuation $p'$ that we can tune, and an unknown, constant attenuation $p_0$. At fixed light source intensity, $p'=1$ can be arbitrarily defined, and deviations therefrom can be measured by comparing $\mu_\mathcal{I}$. Let $\mu_0=I_0p_0N$ let be the mean measured amplitude when $p'=1$. Rewriting equation \ref{eq:var_I_nonpoisson} in a more motivating way:

\begin{equation}
\label{eq:final_quadratic_non_poisson}
\sigma_\mathcal{I}^2=\sigma_i^2+p'\left(I_0\mu_0\right)+p'^2\left(I_0^2p_0^2\left(\sigma^2_N-\lambda \right)\right)
\end{equation}

$\sigma^2_I$ is now written explicitly as a polynomial in $p'$. A measurement of the constant term yields the noise $\sigma_i$, and the linear coefficient, along with the readily measurable $\mu_0$, yields the calibration constant $I_0$. The quadratic coefficient contains all of the terms associated with the non-Poisson emission. Moreover, at constant light-source intensity, this coefficient is constant. Our methodology should therefore be to scan in attenuation at fixed light source intensity, rather than the more traditional scan in intensity at fixed attenuation. Since $\mu_0$ is defined by the user, a fit of $\sigma^2_\mathcal{I}$ to a second order polynomial allows the user to calculate $I_0$ unambiguously.

It can be shown that this result applies also to the more realistic model described in section IIB: retaining our earlier definitions, the relationship will be given as:

\begin{equation}
\label{eq:final_non_poisson_delta}
\sigma_\mathcal{I}^2=\sigma_i^2+p'\left(I_0\left(1+\delta\right)\mu_0\right)+\mathcal{O}(p'^2)
\end{equation}


\bibliography{apssamp}

@misc{qpd2024,
      title={Quantum Parity Detectors: a qubit based particle detection scheme with meV thresholds for rare-event searches}, 
      author={Karthik Ramanathan and John E. Parker and Lalit M. Joshi and Andrew D. Beyer and Pierre M. Echternach and Serge Rosenblum and Brandon J. Sandoval and Sunil R. Golwala},
      year={2024},
      eprint={2405.17192},
      archivePrefix={arXiv},
      primaryClass={physics.ins-det},
      url={https://arxiv.org/abs/2405.17192}, 
}

@misc{celi2026measuringquasiparticledynamicsparticle,
      title={Measuring quasiparticle dynamics for particle impact reconstruction in a superconducting qubit chip}, 
      author={E. Celi and R. Linehan and P. M. Harrington and others},
      year={2026},
      eprint={2604.13176},
      archivePrefix={arXiv},
      primaryClass={quant-ph},
      url={https://arxiv.org/abs/2604.13176}, 
}

@article{Fanofactor,
  title = {Ionization Yield of Radiations. II. The Fluctuations of the Number of Ions},
  author = {Fano, U.},
  journal = {Phys. Rev.},
  volume = {72},
  issue = {1},
  pages = {26--29},
  numpages = {0},
  year = {1947},
  month = {Jul},
  publisher = {American Physical Society},
  doi = {10.1103/PhysRev.72.26},
  url = {https://link.aps.org/doi/10.1103/PhysRev.72.26}
}

@article{irwinQuasiparticleTrapAssisted1995,
  title = {A Quasiparticle-trap-assisted Transition-edge Sensor for Phonon-mediated Particle Detection},
  author = {Irwin, K. D. and Nam, S. W. and Cabrera, B. and others},
  year = {1995},
  month = nov,
  journal = {Review of Scientific Instruments},
  volume = {66},
  number = {11},
  pages = {5322--5326},
  issn = {0034-6748, 1089-7623},
  doi = {10.1063/1.1146105},
  urldate = {2023-02-09},
  langid = {english}
}

@misc{SQUATPaper,
  doi = {10.48550/ARXIV.2310.01345},
  url = {https://arxiv.org/abs/2310.01345},
  author = {Fink,  Caleb W. and Salemi,  Chiara and Young,  Betty A. and Schuster,  David I. and Kurinsky,  Noah A.},
  title = {The Superconducting Quasiparticle-Amplifying Transmon: A Qubit-Based Sensor for meV Scale Phonons and Single THz Photons},
  publisher = {arXiv},
  year = {2023},
  copyright = {Creative Commons Attribution 4.0 International}
}

@article{TwoChannelPaper,
    author = {Anthony-Petersen, R. and Chang, C. L. and Chang, Y.-Y. and others},
    title = {Low energy backgrounds and excess noise in a two-channel low-threshold calorimeter},
    journal = {Applied Physics Letters},
    volume = {126},
    number = {10},
    collaboration={TESSERACT Collaboration},
    pages = {102601},
    year = {2025},
    month = {03},
    issn = {0003-6951},
    doi = {10.1063/5.0247343},
    url = {https://doi.org/10.1063/5.0247343},
    eprint = {https://pubs.aip.org/aip/apl/article-pdf/doi/10.1063/5.0247343/20436183/102601\_1\_5.0247343.pdf},
}

@article{TwoChannelLimits,
  title = {First Limits on Light Dark Matter Interactions in a Low Threshold Two-Channel Athermal Phonon Detector from the {TESSERACT} Collaboration},
  volume = {135},
  ISSN = {1079-7114},
  url = {http://dx.doi.org/10.1103/hsrl-crvf},
  DOI = {10.1103/hsrl-crvf},
  number = {16},
  journal = {Physical Review Letters},
  publisher = {American Physical Society (APS)},
  author = {Bui,  T.K. and Chang,  C.L. and Chang,  Y.-Y. and others},
  year = {2025},
  month = oct 
}

@article{TemplesKID,
  title = {Performance of a phonon-mediated kinetic inductance detector at the NEXUS cryogenic facility},
  volume = {22},
  ISSN = {2331-7019},
  url = {http://dx.doi.org/10.1103/PhysRevApplied.22.044045},
  DOI = {10.1103/physrevapplied.22.044045},
  number = {4},
  journal = {Physical Review Applied},
  publisher = {American Physical Society (APS)},
  author = {Temples,  Dylan J. and Wen,  Osmond and Ramanathan,  Karthik and others},
  year = {2024},
  month = oct 
}

@misc{HabitableWorlds,
      title={Early Architecture Concepts for the Habitable Worlds Observatory -- System Design, Modeling, and Analysis}, 
      author={Alice Liu and Marie Levine and Charley Noecker and others},
      year={2026},
      eprint={2602.11046},
      archivePrefix={arXiv},
      primaryClass={astro-ph.IM},
      url={https://arxiv.org/abs/2602.11046}, 
}

@article{SouthPoleTES,
  author       = {Anderson, A. J. and Ade, P. A. R. and Ahmed, Z. and others},
  title        = {SPT-3G: {A} Multichroic Receiver for the South Pole Telescope},
  journal      = {Journal of Low Temperature Physics},
  volume       = {193},
  number       = {5-6},
  pages        = {1057--1065},
  year         = {2018},
  doi          = {10.1007/s10909-018-2007-z},
  publisher    = {Springer},
}

@article{ATHENA_TES,
  author = {Pajot, F. and Barret, D. and Lam-Trong, T. and den Herder, J.-W. and Piro, L. and Cappi, M. and Huovelin, J. and Kelley, R. and et al.},
  title = {The Athena X-ray Integral Field Unit (X-IFU)},
  journal = {Journal of Low Temperature Physics},
  volume = {193},
  number = {5--6},
  pages = {901--907},
  year = {2018},
  doi = {10.1007/s10909-018-1904-5}
}

@article{BullKID_orig,
    author = {Cruciani, A. and Bandiera, L. and Calvo, M. and others},
    title = {BULLKID: Monolithic array of particle absorbers sensed by kinetic inductance detectors},
    journal = {Applied Physics Letters},
    volume = {121},
    number = {21},
    pages = {213504},
    year = {2022},
    month = {11},
    issn = {0003-6951},
    doi = {10.1063/5.0128723},
    url = {https://doi.org/10.1063/5.0128723},
    eprint = {https://pubs.aip.org/aip/apl/article-pdf/doi/10.1063/5.0128723/16487869/213504\_1\_online.pdf},
}

@article{Calder_Final,
  author       = {Cardani, L. and Casali, N. and Colantoni, I. and others},
  title        = {Final results of CALDER: kinetic inductance light detectors to search for rare events},
  journal      = {The European Physical Journal C},
  volume       = {81},
  pages        = {636},
  year         = {2021},
  doi          = {10.1140/epjc/s10052-021-09454-5},
  url          = {https://doi.org/10.1140/epjc/s10052-021-09454-5},
}

@misc{CDMS_HVeV_2026,
      title={Performance of a SuperCDMS HVeV Detector with Sub-eV Energy Resolution and Single Charge-sensitivity}, 
      author={Kyle Kennard and Aditi Pradeep and Mason Buchanan and others},
      year={2026},
      eprint={2601.16307},
      archivePrefix={arXiv},
      primaryClass={physics.ins-det},
      url={https://arxiv.org/abs/2601.16307}, 
}

@misc{WilliamsFastNeutron,
      title={Low Energy Phonon Bursts Created By Fast Neutron Damage}, 
      author={A. Armatol and C. Augier and L. Bergé and others},
      year={2026},
      collaboration={TESSERACT Collaboration},
      eprint={2603.17964},
      archivePrefix={arXiv},
      primaryClass={physics.ins-det},
      url={https://arxiv.org/abs/2603.17964}, 
}

@article{LANTERN,
title = {LANTERN: A multichannel light calibration system for cryogenic detectors},
journal = {Nuclear Instruments and Methods in Physics Research Section A: Accelerators, Spectrometers, Detectors and Associated Equipment},
volume = {1068},
pages = {169728},
year = {2024},
issn = {0168-9002},
doi = {https://doi.org/10.1016/j.nima.2024.169728},
url = {https://www.sciencedirect.com/science/article/pii/S0168900224006545},
author = {G. {Del Castello}}
}

@book{srivastava1990phonons,
  author    = {G. P. Srivastava},
  title     = {The Physics of Phonons},
  publisher = {Taylor \& Francis},
  year      = {1990},
  address   = {Bristol, UK},
  edition   = {2nd},
  isbn      = {9780367685263}
}

@misc{Nucleus2Channel,
      title={Sensitivity enhancement techniques for cryogenic calorimeters in the NUCLEUS experiment}, 
      author={M. Cappelli and A. Wallach and H. Abele and others},
      year={2026},
      eprint={2603.28276},
      archivePrefix={arXiv},
      primaryClass={physics.ins-det},
      url={https://arxiv.org/abs/2603.28276}, 
}

@article{SiBallCutoff,
  title = {Phonon production in weakly photoexcited semiconductors: Quasidiffusion in Ge, GaAs, and Si},
  author = {Msall, M. E. and Wolfe, J. P.},
  journal = {Phys. Rev. B},
  volume = {56},
  issue = {15},
  pages = {9557--9564},
  numpages = {0},
  year = {1997},
  month = {Oct},
  publisher = {American Physical Society},
  doi = {10.1103/PhysRevB.56.9557},
  url = {https://link.aps.org/doi/10.1103/PhysRevB.56.9557}
}

@article{KIPM_update,
   title={Development Status of the KIPM Detector Consortium},
   ISSN={2378-7074},
   url={http://dx.doi.org/10.1109/TASC.2026.3680780},
   DOI={10.1109/tasc.2026.3680780},
   journal={IEEE Transactions on Applied Superconductivity},
   publisher={Institute of Electrical and Electronics Engineers (IEEE)},
   author={Temples, Dylan J and Smith, Zoë J. and Dang, Selby Q and others},
   year={2026},
   pages={1–9} }

@article{G4CMP,
title = {G4CMP: Condensed matter physics simulation using the Geant4 toolkit},
journal = {Nuclear Instruments and Methods in Physics Research Section A: Accelerators, Spectrometers, Detectors and Associated Equipment},
volume = {1055},
pages = {168473},
year = {2023},
issn = {0168-9002},
doi = {https://doi.org/10.1016/j.nima.2023.168473},
url = {https://www.sciencedirect.com/science/article/pii/S0168900223004631},
author = {M.H. Kelsey and R. Agnese and Y.F. Alam and others}
}

@article{G4CMP_novel,
title = {Modeling athermal phonons in novel materials using the G4CMP simulation toolkit},
journal = {Nuclear Instruments and Methods in Physics Research Section A: Accelerators, Spectrometers, Detectors and Associated Equipment},
volume = {1073},
pages = {170172},
year = {2025},
issn = {0168-9002},
doi = {https://doi.org/10.1016/j.nima.2024.170172},
url = {https://www.sciencedirect.com/science/article/pii/S0168900224010982},
author = {I. Hernandez and R. Linehan and R. Khatiwada and others}
}

@article{BullKID_calibration,
  author    = {Folcarelli, M. and Delicato, D. and Acevedo-Rentería, A. and others},
  title     = {Energy calibration of bulk events in the BULLKID detector},
  journal   = {European Physical Journal C},
  volume    = {86},
  pages     = {301},
  year      = {2026},
  doi       = {10.1140/epjc/s10052-026-15523-4}
}

@article{Stifter_MEMS,
  author = {Stifter, Kelly and Magoon, Hannah and Anderson, Adam J. and others},
  title = {Cryogenic optical beam steering for superconducting device calibration},
  journal = {Journal of Optical Microsystems},
  volume = {5},
  number = {2},
  pages = {024503},
  year = {2025},
  doi = {10.1117/1.JOM.5.2.024503}
}

@article{Kurinsky_MEMS,
author = {Tabassum, N. and Aralis, T. and Anczarski, Jadyn and others},
year = {2025},
month = {10},
pages = {},
title = {Broadband optical modulation and control at millikelvin temperatures},
volume = {96},
journal = {Review of Scientific Instruments},
doi = {10.1063/5.0287214}
}

@phdthesis{Wen2025,
  author       = {Wen, Osmond Zichen},
  title        = {Strategic Planning and Sensitivity-Enhancing Tactics for Detecting Low-Mass Particle Dark Matter with Phonon-Mediated Detectors},
  school       = {California Institute of Technology},
  year         = {2025},
  type         = {PhD thesis},
  doi          = {10.7907/g23k-c067},
  url          = {https://resolver.caltech.edu/CaltechTHESIS:12182024-174539067},
  note         = {ProQuest Dissertations \& Theses, No. 31902836}
}

@article{Cardani_2018,
   title={Al/Ti/Al phonon-mediated KIDs for UV–vis light detection over large areas},
   volume={31},
   ISSN={1361-6668},
   url={http://dx.doi.org/10.1088/1361-6668/aac1d4},
   DOI={10.1088/1361-6668/aac1d4},
   number={7},
   journal={Superconductor Science and Technology},
   publisher={IOP Publishing},
   author={Cardani, L and Casali, N and Cruciani, A and others},
   year={2018},
   month=May, pages={075002} }

\end{document}